\documentclass[nofootinbib, superscriptaddress, twocolumn,secnumarabic,amssymb, nobibnotes, aps, prl, longbibliography]{revtex4-1}
\newcommand{\ket}[1]{|#1\rangle}
\newcommand{\bra}[1]{\langle#1|}
\usepackage{siunitx}
\usepackage{graphicx}
\usepackage{amsmath}
\usepackage{bm}
\usepackage{dsfont} 
\usepackage{color}
\usepackage{isomath}
\usepackage{enumitem}
\usepackage{qcircuit}
\usepackage{hhline}
\usepackage{epsfig}
\usepackage[normalem]{ulem}

\begin{document}

\title{Parallel implementation of high-fidelity multi-qubit gates with neutral atoms}

\newcommand{\Harvard}{Department of Physics, Harvard University, Cambridge, MA
02138, USA}
\newcommand{\Gordon}{Department of Physics, Gordon College, Wenham, MA 01984, USA}
\newcommand{\Chicago}{Pritzker School of Molecular Engineering, University of
Chicago, Chicago, IL 60637, USA}
\newcommand{\ITAMP}{ITAMP, Harvard-Smithsonian Center for Astrophysics,
Cambridge, MA 02138, USA}
\newcommand{\Caltech}{Division of Physics, Mathematics and Astronomy,
California Institute of Technology, Pasadena, CA 91125, USA}
\newcommand{\MIT}{Department of Physics and Research Laboratory of Electronics,
Massachusetts Institute of Technology, Cambridge, MA 02139, USA}

\author{H. Levine}
\affiliation{\Harvard}

\author{A. Keesling}
\affiliation{\Harvard}

\author{G. Semeghini}
\affiliation{\Harvard}

\author{A. Omran}
\affiliation{\Harvard}

\author{T. T. Wang}
\affiliation{\Harvard}
\affiliation{\Gordon}

\author{S. Ebadi}
\affiliation{\Harvard}

\author{H. Bernien}
\affiliation{\Chicago}

\author{M. Greiner}
\affiliation{\Harvard}

\author{V. Vuleti\'c}
\affiliation{\MIT}

\author{H. Pichler}
\affiliation{\Harvard}
\affiliation{\ITAMP}

\author{M. D. Lukin}
\affiliation{\Harvard}

\begin{abstract}
  We report the implementation of universal two- and three-qubit entangling gates on neutral atom qubits encoded in long-lived hyperfine ground states. The gates are mediated by excitation to strongly interacting Rydberg states, and are implemented in parallel on several clusters of atoms  in a one-dimensional array of optical tweezers.
  Specifically, we realize the controlled-phase gate, enacted by a novel, fast protocol  involving only global coupling of two qubits to Rydberg states. We benchmark this operation by preparing Bell states with fidelity $\mathcal{F} \ge 95.0(2)\%$, and extract 
 gate fidelity $\ge 97.4(3)\%$, averaged across five atom pairs.
In addition, we report a proof-of-principle implementation of the three-qubit Toffoli gate, in which two control atoms simultaneously constrain the behavior of one target atom. 
 These experiments demonstrate key ingredients for high-fidelity quantum information processing in a scalable neutral atom platform.
\end{abstract}

\maketitle

Trapped neutral atoms are attractive building blocks for large scale quantum information systems. 
They can be readily manipulated in large numbers while maintaining excellent quantum coherence, as has been demonstrated in remarkable quantum simulation and precision measurement experiments \cite{OpticalLattices2017, PrecisionMeasurementReview2016}.
Single atom initialization, addressing, and readout have been demonstrated in a variety of optical trapping platforms, and single-qubit gates have been implemented with exquisite fidelity \cite{SaffmanSingleQubitGates2015, WeissAddressing2016, WeissSternGerlach2019}. Multi-qubit entangling gates with neutral atoms can be implemented by driving atoms to highly excited Rydberg states, which exhibit strong and long-range interactions~\cite{AntoineReview2016}.
Protocols for entangling atoms using Rydberg interactions have been 
 explored theoretically and experimentally over the last decade \cite{Jaksch2000, SaffmanReview2010, AntoineEntanglementUsingBlockade2010, SaffmanCNOT2010, Biedermann2016, ZhanDifferentIsotopes2017, Pritchard2018}, but despite major advances,  progress in this field has been limited by relatively low fidelities associated with 
 ground-Rydberg state coherent control \cite{SaffmanReview2016}. Recent advances in Rydberg atom control \cite{AntoineCoherence2018, AtomArrayPRL2018, AtomArrayCats2019} offer new opportunities for realization of entangling
 gates, combining 
 high-fidelity performance and  parallelization.

In this Letter, we introduce a new method for realizing multi-qubit entangling gates between individual neutral atoms trapped in optical tweezers. In our approach, qubits are encoded in 
long-lived hyperfine states $\ket{0}$ and $\ket{1}$ which can be coherently manipulated individually or globally to perform single-qubit gates.
Our two-qubit gate, the controlled-phase gate, is implemented with a novel protocol consisting of just two global laser pulses which drive nearby atoms within the Rydberg blockade regime~\cite{Jaksch2000}.
We benchmark this gate by preparing Bell states of two atoms with a fidelity $\mathcal{F} \ge 95.0(2)\%$, averaged across five pairs of atoms. After accounting for state preparation and measurement errors, we extract the entanglement operation fidelity to be $\mathcal{F}^{\rm{c}} \ge 97.4(3)\%$, 
competitive with other leading 
platforms capable of simultaneous manipulation of ten or more qubits \cite{Monroe11Qubits2019, BlattBenchmarking2019, JianWeiPanClusterState2019,IBMGHZ2019}.
We additionally demonstrate a proof-of-principle implementation of the three-qubit Toffoli gate, wherein two atoms simultaneously constrain a third atom through the Rydberg blockade, highlighting the potential use of Rydberg interactions for efficient multi-qubit operations \cite{InnsbruckRydbergMultiqubitGates2010, SaffmanReview2016}.

\begin{figure}
  \includegraphics{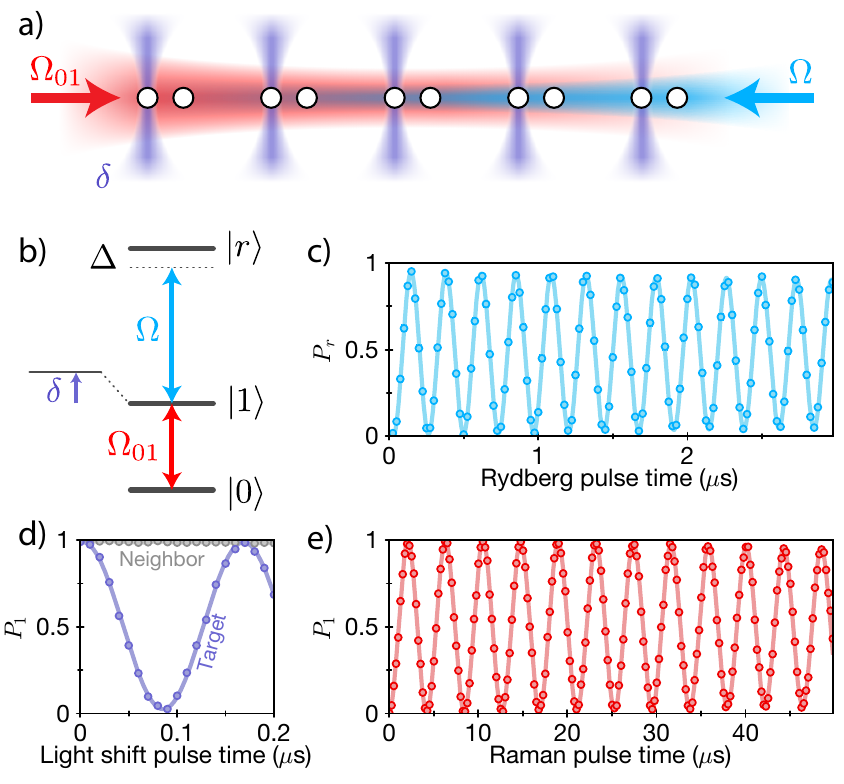}
\caption{\textbf{Control of individual qubits in atom arrays.}
  a) Atoms arranged in pairs are globally driven with a 795~nm Raman laser (shown in red) which couples the hyperfine qubit levels. Local 420~nm beams (purple) are focused onto individual sites, resulting in a light shift $\delta$ used for individual addressing. Additionally, atoms are globally excited by a bichromatic Rydberg laser (shown in blue) from the $\ket{1}$ qubit state to $\ket{r}$. 
b) Relevant atomic levels. The qubit states are $\ket{0} = \ket{5S_{1/2}, F=1, m_F=0}$ and $\ket{1} = \ket{5S_{1/2}, F=2, m_F=0}$. The qubit state $\ket{1}$ is coupled to the Rydberg state $\ket{r} = \ket{70S_{1/2}, m_J=-1/2}$ with detuning $\Delta$ and Rydberg Rabi frequency $\Omega$. 
c) Rydberg Rabi oscillations 
from $\ket{1}$ to $\ket{r}$. Only one atom in each pair is prepared in state $\ket{1}$ to avoid interactions.
d) Local phase shifts as measured in a Ramsey sequence, averaged across the five atom pairs. The purple curve belongs to the addressed atom and shows high-contrast oscillations; the gray curve shows the non-addressed atom, which sees limited $< 2\%$ crosstalk. 
e) Rabi oscillations from $\ket{0}$ to $\ket{1}$ driven by Raman lasers.
Error bars in all figures denote $68\%$ confidence intervals and in most cases are smaller than the markers.
}
\end{figure}

In our experiments, individual atoms are trapped in optical tweezers and sorted by a real-time feedback procedure into groups of two or three, organized in a one-dimensional array \cite{AtomArrayScience2016, AntoineAssembly2016, AhnAssembly2016}. We encode qubits in the hyperfine ground states of these atoms, with $\ket{0} = \ket{5S_{1/2}, F=1, m_F=0}$ and $\ket{1} = \ket{5S_{1/2}, F=2, m_F=0}$. In each experiment we initialize all qubits in $\ket{0}$ through a Raman-assisted optical pumping procedure \cite{Supplement}.
Single-qubit coherent control is achieved through a combination of a global laser field which homogeneously drives all qubits, as well as local addressing lasers which apply AC Stark shifts on individual qubits (Fig. 1a, b). 
The global drive field is generated by a 795~nm laser, tuned near the $5S_{1/2}$ to $5P_{1/2}$  transition. This laser is intensity modulated to produce sidebands which drive the qubits through a two-photon Raman transition with an effective Rabi frequency $\Omega_{01} \approx 2\pi \times 250~$kHz (Fig. 1e) \cite{FastGroundStateManipulation2006, Supplement}. The local addressing beams are generated by an acousto-optic deflector which splits a single 420~nm laser, tuned near the  $5S_{1/2}$ to $6P_{3/2}$ transition, into several beams focused onto individual atoms (Fig.~1a,d) \cite{AtomArrayCats2019}.
We describe these two couplings as global $X(\theta)$ qubit rotations and local $Z(\theta)$ rotations.
After each sequence, we measure the individual qubit states by pushing atoms in $\ket{1}$ out of the traps with a resonant laser pulse, followed by a site-resolved fluorescence image of the remaining atoms \cite{Supplement}.

We perform multi-qubit gates by exciting atoms from the qubit state $\ket{1}$ to the Rydberg state $\ket{r} = \ket{70S_{1/2}, m_J=-1/2}$.
All atoms are homogeneously coupled from $\ket{1}$ to $\ket{r}$ through a two-photon process with effective Rabi frequency $\Omega \approx 2\pi \times 3.5~$MHz (Fig. 1c) \cite{Supplement}. Within a given cluster of atoms, the Rydberg interaction between nearest neighbors is $2\pi \times 24~\rm{MHz} \gg \Omega$; neighboring atoms therefore evolve according to the Rydberg blockade in which they cannot be simultaneously excited to the Rydberg state~\cite{Jaksch2000}.

\begin{figure}
  \includegraphics{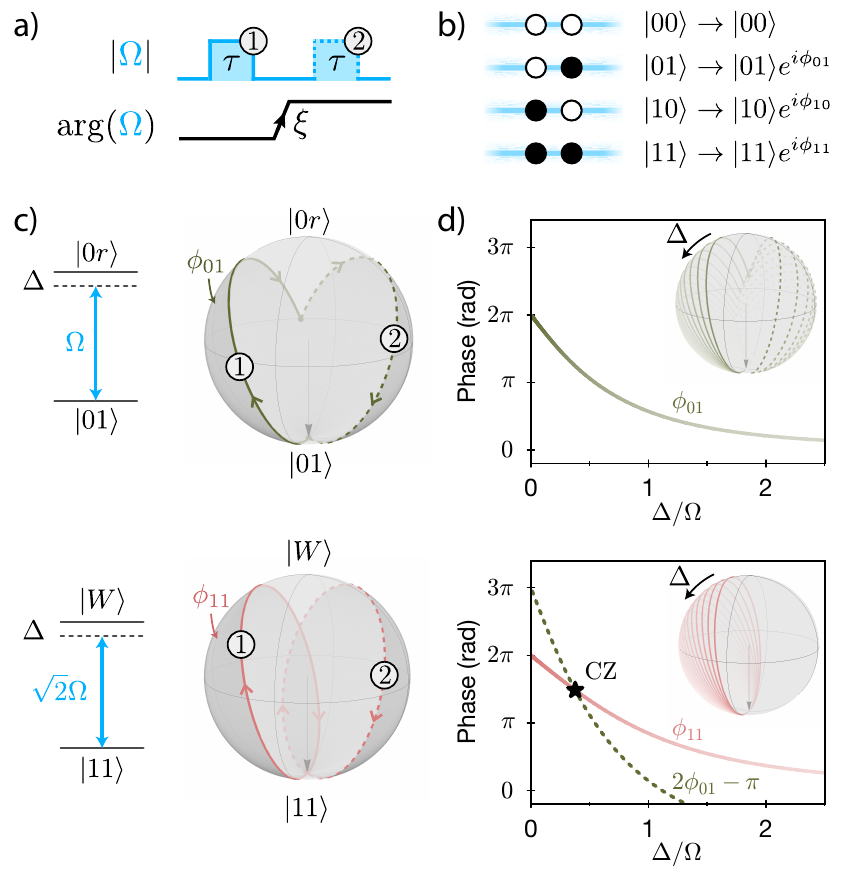}
  \caption{\textbf{Controlled-phase (CZ) gate protocol.} 
	a) 
Two global Rydberg pulses of length $\tau$ and detuning $\Delta$ drive Bloch sphere rotations around two different axes due to a laser phase change $\xi$ between pulses.
b) As a result of the evolution,  each basis state returns to itself with an accumulated dynamical phase. $\ket{00}$ is uncoupled and therefore accumulates no phase. $\ket{01}$ and $\ket{10}$ are equivalent by symmetry ($\phi_{01} = \phi_{10}$), while $\ket{11}$ accumulates phase $\phi_{11}$. The CZ gate is realized for $\phi_{11} = 2\phi_{01} - \pi$.
	c) The dynamics of the $\ket{01}$ and $\ket{11}$ states can be understood in terms of two-level systems with the same detuning $\Delta$ but different effective Rabi frequencies. The pulse length $\tau$ is chosen such that the $\ket{11}$ system undergoes a complete detuned Rabi cycle during the first pulse, while the $\ket{01}$ system undergoes an incomplete oscillation. The laser phase $\xi$ is chosen such that the second pulse drives around a different axis to close the trajectory for the $\ket{01}$ system, while driving a second complete cycle for the $\ket{11}$ system.
	d) The dynamical phases $\phi_{01}$ and $\phi_{11}$  are determined by the shaded area enclosed by the Bloch sphere trajectory and  vary from $2\pi$ to $0$ as a function of  $\Delta$, corresponding to increasingly shallow trajectories. Insets show family of trajectories for different detunings. Choosing $\Delta \approx 0.377 \Omega$ realizes the CZ gate.
}
\end{figure}

To entangle atoms in such arrays, we introduce a new protocol for the two-qubit controlled-phase (CZ) gate that relies only on global excitation of atoms within the Rydberg blockade regime. 
The desired unitary operation CZ maps the computational basis states as follows:
\begin{eqnarray}
  \ket{00} &\to& \ket{00} \nonumber \\
  \ket{01} &\to& \ket{01}e^{i\phi} \nonumber \\
  \ket{10} &\to& \ket{10}e^{i\phi} \nonumber \\
  \ket{11} &\to& \ket{11}e^{i(2\phi - \pi)}
  \label{eq:P_unitary}
\end{eqnarray}
This map is equivalent to the canonical form of the controlled-phase gate
$\mathcal{CZ} = 2\ket{00}\bra{00} - \mathds{1}$
up to a single-qubit phase $\phi$. To realize this gate, we use two global Rydberg laser pulses of the same length $\tau$ and detuning $\Delta$ which couple $\ket{1}$ to $\ket{r}$, with the laser phase of the second pulse shifted by $\xi$ (Fig.~2).

The gate can be understood by considering the behavior of the four computational basis states.
  The $\ket{00}$ state is uncoupled by the laser field and therefore does not evolve.
  The dynamics of $\ket{01}$ (and $\ket{10}$) are given by the coupling of the single atom on the $\ket{1} \leftrightarrow \ket{r}$ transition, forming a two-level system with Rabi frequency $\Omega$ and detuning $\Delta$ (Fig.~2c, top).
The $\ket{11}$ state evolves within the Rydberg blockade regime as a two-level system due to the collective coupling from $\ket{11} \leftrightarrow \ket{W} = \frac{1}{\sqrt{2}}(\ket{1r} + \ket{r1})$, with enhanced Rabi frequency $\sqrt{2}\Omega$ and the same detuning $\Delta$ (Fig.~2c, bottom).
For a chosen detuning $\Delta$, we select the pulse length $\tau$ such that the first laser pulse completes a full cycle of a detuned Rabi oscillation for the $\ket{11}$ system.
The same pulse drives an incomplete Rabi oscillation on the $\ket{01}$ system. A subsequent phase jump $\Omega \to \Omega e^{i\xi}$ rotates the orientation of the drive field around the $Z$ axis by an angle $\xi$ such that a second pulse of length $\tau$ completes the oscillation and returns the state to $\ket{01}$, while driving a second complete detuned oscillation on the $\ket{11}$ configuration. By the end of the second pulse, both $\ket{01}$ and $\ket{11}$ return to their initial positions on the Bloch sphere but with accumulated dynamical phases $\phi_{01}$ and $\phi_{11}$, which depend on the geometric surface area of the Bloch sphere enclosed by the $\Delta$-dependent trajectories. As shown in Fig.~2d, 
for a specific choice of laser detuning ($\Delta \approx 0.377 \Omega$), $2\phi_{01} - \pi = \phi_{11}$, 
realizing the CZ gate \eqref{eq:P_unitary}.
Remarkably, this gate protocol is faster (total time $2\tau \approx 2.732\pi / \Omega$) than the traditional approach \cite{Jaksch2000} of sequential local pulses (total time $4\pi / \Omega$), and offers the additional advantage of requiring only global coupling of both qubits.

\begin{figure}
  \includegraphics{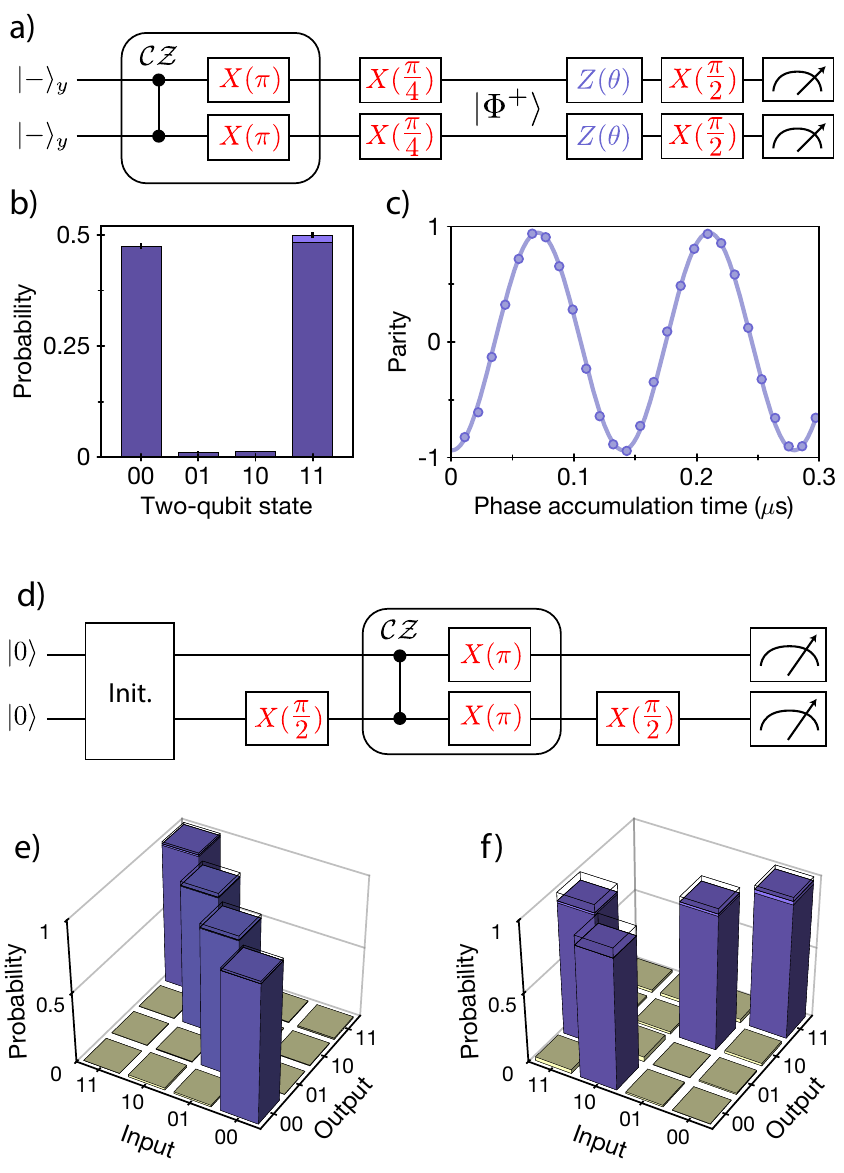}
  \caption{\textbf{Bell state preparation and CNOT gate.}
  a) Quantum circuit used to prepare and probe the $\ket{\Phi^+}$ state. 
  b) Measured populations of the Bell states. Raw measurements associating $\ket{0}$ with atom presence and $\ket{1}$ with atom absence yields $97.6(2)\%$ in the target states. Separate measurements of leakage out of the qubit subspace indicate a small contribution (light shaded region) to these probabilities; subtracting this contribution, the measured population is  $\ge 95.8(3)\%$.
  c) The parity oscillation  with respect to accumulated phase $\theta$ has a measured amplitude of $94.2(4)\%$. The resulting lower bound on Bell state fidelity is $\mathcal{F} \ge 95.0(2)\%$ (raw measurements yield $\mathcal{F}^{\rm{raw}} = 95.9(2)\%$). Correction for SPAM errors results in  $\mathcal{F}^{\rm{c}} \ge 97.4(3)\%$.
  d) The CNOT gate is constructed from our native $\mathcal{CZ}$ gate with the addition of local hyperfine qubit rotations.
  e) The four computational basis states are prepared with average fidelity $96.8(2)\%$.
  f) We apply the CNOT sequence to the four computational basis states and measure the truth table fidelity to be $\mathcal{F}_{\rm{CNOT}} \ge 94.1(2)\%$. Corrected for SPAM errors, the fidelity is $\mathcal{F}^{\rm{c}}_{\rm{CNOT}} \ge 96.5(3)\%$.
  ~Wireframes on purple bars show ideal outcomes; solid bars show the raw measurement;
  the light-shaded top portions of the bars bound the contribution from qubit leakage.  Only the darker lower region is counted towards fidelities.
}
\end{figure}

We demonstrate the parallel operation of the CZ gate on five separate pairs of atoms by using it to create Bell states of the form $\ket{\Phi^+} = \frac{1}{\sqrt{2}}(\ket{00} + \ket{11})$. We initialize all atomic qubits in $\ket{0}$, then apply a global $X(\pi/2)$ Raman pulse to prepare each atom in $\ket{-}_y = \frac{1}{\sqrt{2}}(\ket{0} - i \ket{1})$.
The CZ gate protocol, consisting of the two Rydberg laser pulses, is then applied over a total time of $0.4~\mu$s, during which the optical tweezers are turned off to avoid anti-trapping of the Rydberg state. The pulse sequence realizes map $\eqref{eq:P_unitary}$, along with an additional phase rotation on each qubit due to the light shift of the Rydberg lasers on the hyperfine qubit states. We embed the CZ implementation in an echo sequence to cancel the effect of the light shift, and we add an additional short light shift to eliminate the single-particle phase $\phi$ \cite{Supplement}. Altogether, this realizes a unitary that combines the canonical $\mathcal{CZ}$ gate with a global $X(\pi)$ gate (enclosed region in Fig. 3a,d). A final $X(\pi/4)$ rotation produces the Bell state $\ket{\Phi^+}$ (Fig. 3a) \cite{Supplement}.

We characterize the experimentally produced state $\rho$ by evaluating its fidelity with respect to the target Bell state $\mathcal{F} = \langle \Phi^+ | \rho | \Phi^+\rangle$. The fidelity is the sum of two terms, the first of which is the Bell state populations, given by the probability of observing $\ket{00}$ or $\ket{11}$  (Fig.~3b). The second term  is the coherence between $\ket{00}$ and $\ket{11}$, measured by applying a global $Z(\theta)$ rotation followed by a global $X(\pi/2)$ rotation and observing parity oscillations (Fig.~3a,c) \cite{Sackett2000}.
When evaluating the contributions to the fidelity, we account for atom population left in the Rydberg state after the operation and for background losses. Both of these correspond to leakage out of the qubit subspace and can lead to overestimation of the $\ket{1}$ populations  and Bell state fidelities in the raw measurements. Using separate measurements of atoms in both hyperfine qubit states \cite{Supplement}, we determine a conservative upper bound on these leakage errors and 
subtract this contribution (shown in light shaded regions of bar plots in Figs. 3,4, see \cite{Supplement}). The resulting lower bound on the Bell state fidelity is $\mathcal{F} \ge 95.0(2)\%$.

The measured Bell state fidelity includes errors in state preparation and measurement (SPAM), as well as errors in the two-qubit entangling gate. To characterize the entangling gate specifically, we evaluate the error contributions from SPAM ($1.2(1)\%$ per atom) and compute a SPAM-corrected fidelity $\mathcal{F}^{\rm{c}} \ge 97.4(3)\%$ \cite{Supplement}. The majority of the remaining error is due to finite atomic temperature and laser scattering during Rydberg dynamics \cite{Supplement}. 
We separately characterize our native $\mathcal{CZ}$ gate by converting it to a controlled-NOT (CNOT) gate via local rotations (Fig.~3d).
We measure the action of the CNOT gate on each computational basis state to obtain its truth table fidelity $\mathcal{F}^{\rm{c}}_{\rm{CNOT}} \ge 96.5(3)\%$, corrected for SPAM errors (Fig.~3e,f) \cite{Supplement}.

\begin{figure}
  \includegraphics{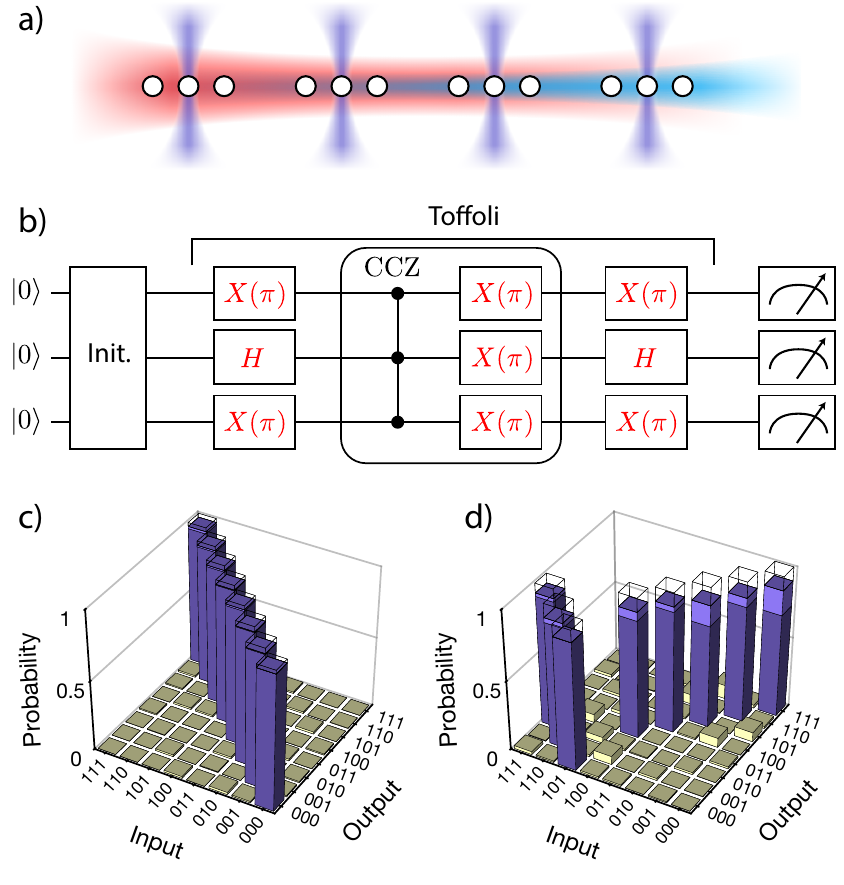}
  \caption{\textbf{Realization of three-qubit Toffoli gate.}
	a) The Toffoli gate is implemented in parallel on four triplets of atomic qubits using the same lasers as for two-qubit gates.
	b) Quantum circuit for constructing the Toffoli gate from local rotations and a globally implemented CCZ gate. 
	c) Eight computational basis states are prepared with average fidelity $95.3(3)\%$.
	d) Measured truth table, with fidelity $\mathcal{F}_{\rm{Toff}} \ge 83.7(3)\%$. Corrected for SPAM errors, the fidelity is $\mathcal{F}_{\rm{Toff}}^{\rm{c}} \ge 87.0(4)\%$.
~Wireframes on purple bars show ideal outcomes; solid bars show the raw measurement;
the light-shaded top portions of the bars bound the contribution from qubit leakage.  Only the darker lower region is counted towards fidelities.
  }
\end{figure}
Finally, we extend our control of multiple atomic qubits to implement the three-qubit controlled-controlled-phase  (CCZ) gate. This
logic operation can be decomposed into five two-qubit gates \cite{ElementaryGates1995, WalraffToffoli2012, MonroeGrover2017}.
Instead, we realize this multiple-control gate directly by preparing three atoms in the nearest-neighbor blockade regime such that both outer atoms constrain the behavior of the middle atom. In this configuration, the CCZ gate can be approximately implemented with an amplitude and frequency modulated global laser pulse \cite{Supplement}, which we numerically optimize through the remote dressed chopped random basis (RedCRAB) optimal control algorithm \cite{dCrab2015, REdCrab2018}.

We implement the CCZ gate in parallel on four triplets of atomic qubits (Fig.~4a). The three atoms in each triplet are arranged such that nearest neighbors are blockaded by the strong $2\pi \times 24~$MHz interaction, as in the two-qubit experiments. The edge atoms interact with each other weakly ($2\pi \times 0.4~$MHz).
As with the two-qubit gate, we embed the CCZ gate in an echo sequence to cancel light shifts, such that our gate implements CCZ along with a global $X(\pi)$ rotation. To characterize the performance of this three-qubit gate, we convert it into a Toffoli gate by applying a local Hadamard on the middle atom before and after the CCZ gate (along with edge $X(\pi)$ pulses, to simplify implementation \cite{Supplement}) (Fig.~4b).
We apply the Toffoli gate to each computational basis state to measure the truth table fidelity $\mathcal{F}^{\rm{c}}_{\rm{Toff}} \ge 87.0(4)\%$, corrected for SPAM errors (Fig.~4c,d) \cite{Supplement}.
We additionally perform ``limited tomography'', consisting of truth table measurements in a rotated basis, to verify the phases of the Toffoli unitary in a more experimentally accessible manner than full process tomography \cite{MonroeGrover2017}. The limited tomography fidelity is $\mathcal{F}^{\rm{c}}_{\rm{LT}} \ge 86.2(6)\%$ \cite{Supplement}.

These results can be directly improved and extended along several directions. 
The fidelity of Rydberg coupling is primarily limited by finite atomic temperature and off-resonant laser scattering, which can be addressed by sideband cooling of atoms within optical tweezers \cite{AdamSidebandCooling2012, JeffSidebandCooling2013} and by higher power lasers. The background atomic loss and state preparation can be improved using higher quality vacuum systems \cite{EndresSrTweezers2019} and more sophisticated state preparation techniques \cite{WeissSternGerlach2019}. Finally, atomic qubit readout can be improved using 
recently demonstrated non-destructive readout protocols \cite{WeissSternGerlach2019, MeschedeReadout2017, SaffmanReadout2017}
to give stronger constraints on the atomic populations.

While in this work we have performed parallel gate implementation on spatially separated clusters of atoms, 
the same approach can be extended to non-local coupling within contiguous atom arrays using local addressing 
with an additional off-resonant laser system.
Specifically, subsets of the array could be simultaneously illuminated 
to create light shifts that bring them into resonance with a global resonant Rydberg excitation laser \cite{Supplement}.
Furthermore, with more atoms arranged in the blockade volume, the controlled-phase gate demonstrated here can be extended to higher multi-qubit gates  with global coupling \cite{Supplement}. The dipolar interaction between $S$ and $P$ Rydberg states \cite{AntoineSP2017} could also be used to achieve improved gate connectivity between qubits. 
A combination of the present results with recently demonstrated 
2D and 3D neutral atom arrays \cite{AntoineAssembly2016, Antoine3D, WeissAssembly2018} 
will be well-suited for the implementation of deep quantum circuits or variational quantum optimization with hundreds of qubits \cite{PichlerMIS2018}.  In addition, such a platform could be utilized to explore efficient methods for error correction and fault-tolerant  operation to eventually enable scalable quantum processing. 

We thank Tommaso Calarco, Simone Montangero, Jian Cui, Marco Rossignolo, and Phila Rembold for the remote use of their RedCRAB optimal control server, and Manuel Endres and Alexander Zibrov for useful discussions and contributions to the experiment.  We  acknowledge financial support from the Center for Ultracold Atoms, the National Science Foundation, Vannevar Bush Faculty Fellowship, the US Department of Energy, and the Office of Naval Research.
H.L.  acknowledges support from the National Defense Science and Engineering Graduate (NDSEG) fellowship.
G.S.  acknowledges support from a fellowship from the Max Planck/Harvard Research Center for Quantum Optics.

\textit{Note added:} During the completion of our manuscript we became aware of related work demonstrating neutral atom gates in two-dimensional atom arrays~\cite{Saffman2019}.

%

\clearpage

\onecolumngrid

\begin{center}
  \textbf{\large Supplementary information}\\[.2cm]
  \vspace{0.2cm}
\end{center}

\twocolumngrid

\renewcommand{\thefigure}{S\arabic{figure}}
\renewcommand{\thetable}{S\arabic{table}}
\setcounter{figure}{0}
\setcounter{equation}{0}

\begin{figure*}
\includegraphics{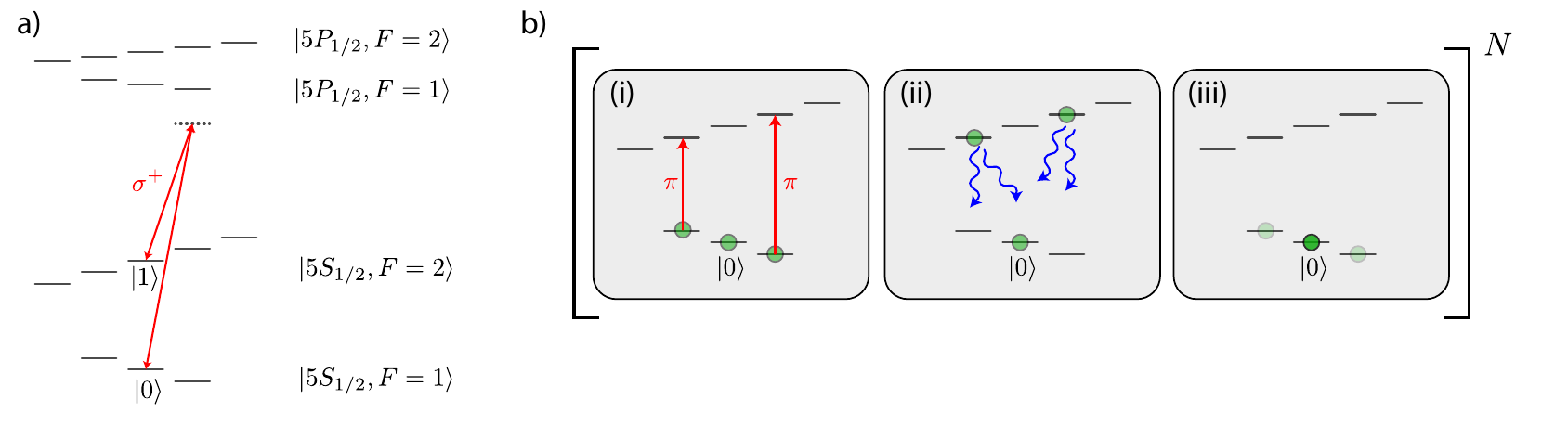}
\caption{\textbf{Raman laser and optical pumping.} a) Level diagram. The Raman laser is bichromatic and contains two frequency components separated by $2\pi \times 6.8~$GHz. These frequencies are red-detuned by $2\pi \times 100~$GHz to the $\ket{5P_{1/2}}$ manifold. b) Raman-assisted optical pumping. (i) We begin by coarse pumping into all three sublevels of $F=1$, and apply a Raman $\pi$-pulse to excite from $\ket{F=1, m_F=-1}$ to $\ket{F=2, m_F=-1}$ and from $\ket{F=1, m_F=+1}$ to $\ket{F=2, m_F=+1}$. (ii) We then coarse pump back from $F=2$ to $F=1$. (iii) The net effect is to transfer some population from $\ket{F=1, m_F=\pm1}$ to $\ket{F=1, m_F=0}$. We repeat this cycle $N=70$ times and achieve a net population of $99.3(1)\%$ in $\ket{0} = \ket{F=1, m_F=0}$.}
\label{fig:raman}
\end{figure*}
\section{Raman laser}

We drive transitions between our qubit states using a 795~nm Raman laser which is $2\pi \times 100~$GHz red-detuned from the $5S_{1/2}$ to $5P_{1/2}$ transition. We couple the laser into a fiber-based Mach-Zehnder intensity modulator (Jenoptik AM785) which is DC biased around minimum transmission. The modulator is driven at half the qubit frequency ($\omega_{01} = 2\pi \times 6.83~$GHz), resulting in sidebands at $\pm 2\pi \times 3.42~$GHz, while the carrier and higher order sidebands are strongly suppressed. This approach is passively stable on the timescale of one day without any active feedback, in contrast with other approaches to generate sidebands through phase modulation and then separate suppression of the carrier mode with free space optical cavities or interferometers.

The Raman laser is aligned along the array of atoms (co-aligned with the $8.5~$G bias magnetic field) and is $\sigma^+$ polarized, such that the two sidebands coherently drive $\pi$ transitions between the $F=1$ and $F=2$ ground state manifolds with a Rabi frequency of $\Omega = 2\pi \times 250~$kHz (Fig.~\ref{fig:raman}a). The Raman drive light induces a differential light shift of $2\pi \times 20~$kHz on the qubit transition; we adjust the drive frequency of the intensity modulator to correct for this light shift when we apply a Raman pulse.

\section{Optical pumping into $\ket{0}$}
We optically pump atoms into $\ket{0} = \ket{5S_{1/2}, F=1, m_F=0}$ using a Raman-assisted pumping scheme with an 8.5 G magnetic field. As illustrated in Fig.~\ref{fig:raman}b, we begin by coarse pumping of atoms into all $m_F$ states within the $\ket{5S_{1/2}, F=1}$ manifold by shining resonant light on the $\ket{5S_{1/2}, F=2}$ to $\ket{5P_{3/2}, F=2}$ transition. We then apply a Raman $\pi$ pulse at a detuning that drives population from $\ket{F=1, m_F=-1}$ to $\ket{F=2, m_F=-1}$. A second pulse drives population from $\ket{F=1, m_F=+1}$ to $\ket{F=2, m_F=+1}$. The process then repeats by again coarse pumping any population that was transferred to $F=2$ back into the $F=1$ manifold. The net effect of one cycle is to transfer a portion of the population in $\ket{F=1, m_F=\pm 1}$ into $\ket{F=1, m_F=0}$. We repeat this cycle 70 times over a duration of $300~\mu$s to achieve a $\ket{0}$ preparation fidelity of $99.3(1)\%$.

\section{Rydberg laser system}
We couple atoms from $\ket{1} = \ket{5S_{1/2}, F=2, m_F=0}$ to $\ket{r} = \ket{70S_{1/2}, m_J=-1/2}$ through a two-color laser system at $420~$nm and $1013~$nm, described in \cite{SIAtomArrayCats2019}. The lasers are polarized to drive $\sigma^{-}$ and $\sigma^+$ transitions, respectively, through the intermediate state $\ket{6P_{3/2}}$.
In previous experiments using $\ket{5S_{1/2}, F=2, m_F=-2}$ as the ground state level, selection rules ensured that only a single intermediate sublevel within $\ket{6P_{3/2}}$ and only a single Rydberg state could be coupled. Additionally, the combined two-photon transition was magnetically insensitive.

Coupling from $\ket{1} = \ket{5S_{1/2}, F=2, m_F=0}$ to Rydberg states, as in these experiments, adds a few complications. Firstly, multiple intermediate states are coupled and both $\ket{70S_{1/2}, m_J= \pm 1/2}$ sublevels within the Rydberg manifold can be reached. This requires working at a higher magnetic field to spectrally separate the $m_J = \pm 1/2$ Rydberg levels. In these experiments, we work at a magnetic field of $8.5~$G such that the splitting between $m_J = \pm 1/2$ is $2\pi \times 24.8~$MHz. The matrix element is also reduced in the coupling from $\ket{1}$ to $\ket{r}$ while the laser scattering rate stays the same; additionally, the transition is now magnetically sensitive. Nonetheless, this scheme benefits from high-quality qubit states $\ket{0}$ and $\ket{1}$ within the ground state manifold which can be easily coupled with a Raman laser system and which preserve coherence in optical tweezers.

\section{Constructing quantum circuits from native single-qubit gates}
All pulse sequences shown in the main text are decomposed into pre-calibrated single-qubit gates (and, where indicated, global multi-qubit gates). The two single-qubit gates are $X(\pi/4)$, implemented globally on all qubits simultaneously, and $Z(\pi)$, implemented by a light shift from a laser focused onto a single atom. In practice, the local $Z(\pi)$ gates are applied to one atom from each cluster at the same time (i.e., the left atom of each cluster or the middle of each cluster).

\subsection{Initializing computational basis states}

For two qubits, we initialize all four computational basis states using global $X(\pi/2)$ pulses (consisting of two sequential $X(\pi/4)$ gates) and local $Z(\pi)$ gates on the left atom only (top qubit in each circuit). The $\ket{00}$ state requires no pulses to prepare, and the $\ket{11}$ state requires only a global $X(\pi)$ gate. We prepare $\ket{01}$ as follows:
\centerline{\Qcircuit @C=0.8em @R=1.2em {
\\
|0\rangle & & \qw  & \gate{X(\pi/2)} & \gate{Z(\pi)} & \gate{X(\pi/2)} & \qw & |0\rangle \\
|0\rangle & & \qw  & \gate{X(\pi/2)} & \qw & \gate{X(\pi/2)} & \qw & |1\rangle \\
\\
}}
and $\ket{10}$ according to

\centerline{\Qcircuit @C=0.8em @R=1.2em {
\\
|0\rangle & & \qw  & \gate{X(\pi/2)} & \gate{Z(\pi)} & \gate{X(3\pi/2)} & \qw & |1\rangle \\
|0\rangle & & \qw  & \gate{X(\pi/2)} & \qw & \gate{X(3\pi/2)} & \qw & |0\rangle \\
\\
}}

For three qubits, we initialize the eight computational basis states again using global $X(\pi/2)$ pulses and local $Z(\pi)$ pulses which can be applied to any of the three atoms. $\ket{000}$ and $\ket{111}$ can again be prepared with either no operation or with a global $X(\pi)$ gate, respectively. Other states have one atom in $\ket{1}$ and the other two in $\ket{0}$, or vice versa. We illustrate how both configurations are prepared by showing two examples. First, $\ket{100}$:

\centerline{\Qcircuit @C=0.8em @R=1.2em {
\\
|0\rangle & & \qw  & \gate{X(\pi/2)} & \gate{Z(\pi)} & \gate{X(3\pi/2)} & \qw & |1\rangle \\
|0\rangle & & \qw  & \gate{X(\pi/2)} & \qw & \gate{X(3\pi/2)} & \qw & |0\rangle \\
|0\rangle & & \qw  & \gate{X(\pi/2)} & \qw & \gate{X(3\pi/2)} & \qw & |0\rangle \\
\\
}}

Next, we consider preparation of $\ket{110}$, which requires instead local addressing on the rightmost atom.

\centerline{\Qcircuit @C=0.8em @R=1.2em {
\\
|0\rangle & & \qw  & \gate{X(\pi/2)} & \qw & \gate{X(\pi/2)} & \qw & |1\rangle \\
|0\rangle & & \qw  & \gate{X(\pi/2)} & \qw & \gate{X(\pi/2)} & \qw & |1\rangle \\
|0\rangle & & \qw  & \gate{X(\pi/2)} & \gate{Z(\pi)} & \gate{X(\pi/2)} & \qw & |0\rangle \\
\\
}}

\subsection{Local $X(\pi/2)$ for CNOT gate}
To convert the $\mathcal{CZ}$ gate to the CNOT gate, we apply a local $X(\pi/2)$ before and after the gate to the target atom. We implement this as follows:
\centerline{\Qcircuit @C=0.8em @R=1.2em {
\\
& & \qw  & \gate{X(\pi/4)} & \gate{Z(\pi)} & \gate{X(\pi/4)} & \qw & = & & \qw & \gate{Z(\pi)} & \qw\\
& & \qw  & \gate{X(\pi/4)} & \qw & \gate{X(\pi/4)} & \qw & = & & \qw & \gate{X(\pi/2)} & \qw\\
\\
}}
This circuit applies a local $X(\pi/2)$ on the right atom; while it additionally applies a $Z(\pi)$ gate on the left atom, this circuit is only applied in a context in which the left atom is in a computational basis state $\ket{0}$ or $\ket{1}$, in which case the $Z(\pi)$ gate only introduces a global phase and therefore plays no role. In general, applying additional $Z(\pi)$ gates could be used to cancel the effect on the left atom, but this was not necessary for these experiments.

\subsection{Local Hadamard for Toffoli implementation}
To convert the CCZ gate to a Toffoli gate, we apply a local rotation on the target (middle) qubit before and after the CCZ pulse. The simplest method to accomplish this given our native gate set is to apply a global $X(\pi/4)$, followed by a local $Z(\pi)$ on the middle qubit, and then a global $X(3\pi/4)$.

\centerline{\Qcircuit @C=0.8em @R=1.2em {
\\
& \qw & \gate{X(\pi/4)} & \qw & \gate{X(3\pi/4)} & \qw & & & & \qw & \gate{X(\pi)} & \qw \\
& \qw & \gate{X(\pi/4)} & \gate{Z(\pi)} & \gate{X(3\pi/4)} & \qw & & = & &\qw & \gate{H} & \qw\\
& \qw & \gate{X(\pi/4)} & \qw & \gate{X(3\pi/4)} & \qw &  & & & \qw & \gate{X(\pi)} & \qw \\
\\
}}

On each edge qubit, the net effect is simply a $X(\pi)$ gate. On the middle qubit, this sequence constitutes a Hadamard gate (defined along a different axis than the typical definition), where
\begin{equation}
  H = \frac{1}{\sqrt{2}}
  \begin{pmatrix}
    1 & i \\
    -i & -1 
  \end{pmatrix}
\end{equation}

\section{Design of two-qubit CZ gate}
In this section we provide a detailed theoretical discussion of the two-qubit gate realized in the experiment. The desired unitary operation maps the computational basis states as follows:
\begin{align}\label{CZ1}
&\ket{00}\rightarrow \ket{00}\nonumber\\
&\ket{01}\rightarrow \ket{01}\nonumber\\
&\ket{10}\rightarrow \ket{10}\nonumber\\
&\ket{11}\rightarrow \ket{11}e^{i\pi}
\end{align}
Up to a global gauge choice (i.e. global rotation of the qubits), this is equivalent to the following gate
\begin{align}\label{CZ2}
&\ket{00}\rightarrow \ket{00}\nonumber\\
&\ket{01}\rightarrow \ket{01}e^{i\phi_1}\nonumber\\
&\ket{10}\rightarrow \ket{10}e^{i\phi_1}\nonumber\\
&\ket{11}\rightarrow \ket{11}e^{i(2\phi_1+\pi)}
\end{align}
where $\phi_1$ is arbitrary. 

To realize such a gate we drive both atoms globally and homogeneously with a laser that couples state $\ket{1}$ to the Rydberg state $\ket{r}$. This can be achieved via a single laser field or by a two-photon process.  The Hamiltonian governing the dynamics of a pair of atoms is given by
\begin{equation*}
H=\sum_{i=1}^2\frac{1}{2}(\Omega \ket{1}_i\bra{r}+\Omega^*\ket{r}_i\bra{1})-\Delta\ket{r}_i\bra{r}+V\ket{r}_1\bra{r}\otimes \ket{r}_2\bra{r}
\end{equation*}
where $\Delta$ is the detuning of the excitation laser from the transition frequency between states $\ket{1}$ and $\ket{r}$, and $\Omega$ is the corresponding Rabi frequency. The interaction strength between two atoms in Rydberg states is given by $V$. In the following analysis we first assume that $V\gg |\Omega|, |\Delta|$, which can be realized by trapping the atoms sufficiently close to each other. This so-called Rydberg-blockade regime simplifies the following discussion, but is not crucial for the realization of the gate. 

The dynamics of the system decouples into a few simple sectors:\\
(i) The state $\ket{00}$ doesn't evolve. \\
(ii) If one of the atoms is in $\ket{0}$, only the other system evolves. The dynamics is thus equivalent to that of a two level system (TLS) with states $\ket{1}=\ket{a_1}$ and $\ket{r}=\ket{b_1}$ and Hamiltonian $$H_1=\frac{1}{2}(\Omega \ket{a_1}\bra{b_1}+\Omega^*\ket{b_1}\bra{a_1})-\Delta\ket{b_1}\bra{b_1}.$$
(iii) If both atoms are initially in state $\ket{1}$, then the dynamics is again equivalent to that of an effective  single TLS, formed by the states $\ket{11}=\ket{a_2}$ and $\frac{1}{\sqrt{2}}(\ket{r,1}+\ket{1,r})=\ket{b_2}$, with Hamiltontian $$H_2=\frac{\sqrt{2}}{2}(\Omega \ket{a_2}\bra{b_2}+ \Omega^*\ket{b_2}\bra{a_2})-\Delta\ket{b_2}\bra{b_2}.$$
This assumes a perfect Rydberg blockade, equivalent to $V\rightarrow \infty$. We stress again that this assumption simplifies the analysis but is not necessary to realize our proposed gate. 

The controlled-phase gate can be constructed from two identical global pulses of the Rydberg laser field, with equal duration $\tau$ and detuning $\Delta$, along with a phase jump by $\xi$ in between. Each pulse changes the state of the atoms according to the unitary ${U}=\exp(-iH\tau)$. The change of the laser phase between pulses, $\Omega \to \Omega e^{i\xi}$, effectively corresponds to driving the system around a different axis on the Bloch sphere.

Let us examine how the four computational basis states evolve under the action of $\mathcal{U}$, which describes the effect of both laser pulses combined. First we note that  $\mathcal{U}\ket{00}=\ket{00}$.  Thus the unitary $\mathcal{U}$ maps the state $\ket{00}$ as expected for the CZ gate. 

Next, let us consider the evolution of state $\ket{11}$. We choose the length of each pulse $\tau$ such that a system prepared in state $\ket{11}$ undergoes a complete, detuned Rabi oscillation and returns to the state $\ket{11}$ already after the first single pulse; that is, $U\ket{11}=e^{i\phi_2/2}\ket{11}$. This is guaranteed by the choice \begin{align}\label{time}\tau=2\pi/\sqrt{\Delta^2+2\Omega^2}.\end{align}  The second pulse also leads to a complete, detuned Rabi cycle about a different axis, but results in the same accumulated phase. In total, we find $\mathcal{U}\ket{11} = e^{i\phi_2} \ket{11}$. The dynamical phase accumulated by this process is given by $\phi_2=2\pi\times 2\Delta /\sqrt{\Delta^2+2\Omega^2}$.

Finally, let us consider the evolution of the states $\ket{01}$ and $\ket{10}$.  In each case, this is also described by a detuned Rabi oscillation. However, due to to the mismatch between the effective Rabi frequencies in $H_1$ and $H_2$, the the state $\ket{10}$ ($\ket{01}$) does not return to itself after the time $\tau$ but a superposition state is created: 
$U\ket{10}=\cos(\alpha)\ket{10}+\sin(\beta)e^{i\gamma}\ket{r0}$, and $U\ket{01}=\cos(\alpha)\ket{01}+\sin(\beta)e^{i\gamma}\ket{0r}$
. The real coefficients $\alpha$, $\beta$ and $\gamma$ are determined by the choice of $\Omega$, $\Delta$ and $\tau$, and can easily be calculated (we omit explicit expressions here for compactness). Crucially, by a proper choice of the phase jump between the two pulses, $\xi$, one can always guarantee that the the system returns to the state $\ket{10}$  ($\ket{01}$) after the second pulse. This can be calculated to be
\begin{align}\label{phase}
e^{-i\xi}=\frac{-\sqrt{y^2+1} \cos \left(\frac{1}{2} s \sqrt{y^2+1}\right)+i y \sin
   \left(\frac{1}{2} s \sqrt{y^2+1}\right)}{\sqrt{y^2+1} \cos \left(\frac{1}{2} s
   \sqrt{y^2+1}\right)+i y \sin \left(\frac{1}{2} s \sqrt{y^2+1}\right)}
\end{align}
where we use the short hand notation $y=\Delta/\Omega$ and $s=\Omega \tau$. With this choice of the phase we thus have $\mathcal{U}\ket{10} = e^{-i \phi_1} \ket{10}$ and $\mathcal{U}\ket{01} = e^{-i \phi_1} \ket{01}$. The acquired dynamical phase can be calculated using straightforward algebra, and is a function of $\Delta/\Omega$, $\tau \Omega$ and $\xi$. Since we fixed $\tau$ in equation \eqref{time}, and $\xi$ in \eqref{phase},  $\phi_1$ is actually solely determined by the dimensionless quantity $\Delta/\Omega$. Note that also $\phi_2$ is only a function of $\Delta/\Omega$. However, the functional dependence is different, and we can find a choice for $\Delta/\Omega$ such that $e^{i \phi_2} = e^{i(2\phi_1 + \pi)}$ (see Fig.~2 of main text). With this choice, we obtain exactly the gate given in \eqref{CZ2} which is equivalent to the controlled-phase gate \eqref{CZ1} (up to trivial single qubit rotations). For completeness we give the corresponding numerical values of the relevant parameters:
\begin{align}\label{CZparams}
&\Delta/\Omega=0.377371\\
&\xi=3.90242\\
&\Omega \tau=4.29268
\end{align}

Finally, we note that this construction can be generalized to multi-qubit controlled phase gates in fully blockaded systems with more than two atoms.

\subsection{Accounting for imperfect blockade}
The above analysis is based on the perfect blockade mechanism. Finite blockade interactions (and other experimental imperfections, such as coupling to other Rydberg states) can be accounted for, and lead only to an effective renormalization of the parameters given in \eqref{CZparams}.
To see this, note that a finite value of $V$ only affects the dynamics if the system is initially in the state $\ket{11}$. Instead of being restricted to the two states $\ket{a_2}=\ket{11}$ and $\ket{b_2}=\ket{1r}+\ket{r1}$, a third state $\ket{c_2}=\ket{rr}$ has to be considered, and $H_2$ is replaced by
\begin{align}
H_2=&\frac{\sqrt{2}}{2}(\Omega \ket{a_2}\bra{b_2}+ \Omega\ket{b_2}\bra{c_2}+\Omega^*\ket{c_2}\bra{b_2}+\Omega^*\ket{b_2}\bra{a_2})\nonumber\\
&-\Delta\ket{b_2}\bra{b_2}+(V-2\Delta)\ket{c_2}\bra{c_2}.
\end{align}
For $V\gg|\Delta|,|\Omega|$, the effect for finite blockade simply reduces to the two-level system $\{\ket{a_2},\ket{b_2}\}$ where $\Delta$ is renormalized by an amount $\Omega^2/(2V)$. 
Even for small $V>0$ and a given $\Delta$, we can always choose $\Omega$ and $\tau$ such that the system initialized in the state $\ket{a_2}$ returns after the first pulse. Thus finite blockade simply replaces the complete Rabi oscillation in the fully blockaded regime, by a slightly more complicated, but still closed path in a two-dimensional Hilbert space. The analysis of the dynamics of the other computational basis states is unaffected by the finite value of $V$. It is thus straightforward to ensure that a system initially in the state $\ket{10}$ returns to $\ket{10}$ for each choice of $V$ and $\Delta$. This allows one to use $\Delta$ as a control knob for the relative dynamical phases acquired by $\ket{11}$ and $\ket{10}$, and thus realize a $\text{CZ}$ gate. 

\section{Experimental calibration of CZ gate}
The CZ gate requires two laser pulses with a relative phase shift between them. The detuning of the two pulses $\Delta$ is determined relative to the experimentally calibrated Rydberg resonance by numerical calculations. The pulse time and the phase jump between pulses both require experimental calibration due to perturbations in timing and phase asssociated with an AOM-based control system. The pulse time $\tau$ is calibrated first by preparing both atoms in the qubit pair in $\ket{1}$ and driving at detuning $\Delta$ to the Rydberg state. We observe detuned Rabi oscillations to the symmetrically excited state $\ket{W} = \frac{1}{\sqrt{2}}(\ket{1r} + \ket{r1})$ and extract the pulse time at which the population returns fully to $\ket{11}$.

After fixing $\tau$, we prepare only single isolated atoms in $\ket{1}$ and we drive two pulses of length $\tau$ with a variable relative phase. By identifying the phase for which the single atom returns fully to $\ket{1}$ by the end of the sequence, we fix the relative phase $\xi$.

Finally, we calibrate the global phase shift necessary to convert the CZ gate (with single-particle phase $\phi$) into the canonical form:
\begin{equation}
  {\mathcal{CZ}} = 
  \begin{pmatrix}
	1 & 0 & 0 & 0 \\ 
	0 & -1 & 0 & 0 \\ 
	0 & 0 & -1 & 0 \\ 
	0 & 0 & 0 &-1
  \end{pmatrix}
\end{equation}

We implement this phase correction by applying the global 420 nm laser for a fixed time in the absence of the 1013 nm Rydberg light; this avoids any resonant Rydberg excitation and instead only adds a phase shift. To calibrate the phase correction, we apply the Bell state sequence in which we attempt to prepare the Bell state $\ket{\Phi^+}$ and then we apply an additional $X(\pi/2)$ rotation to both qubits. If our phase correction is optimal, we should prepare the state $\ket{\Psi^+}$, which we can measure in populations. We vary the global phase correction to maximize the measured populations in $\ket{\Psi^+}$ at the end of this sequence.

\section{Preparation of Bell state using $\mathcal{CZ}$ gate and $\pi/4$ pulse}
Our global implementation of the $\mathcal{CZ}$ gate enables the preparation of Bell states with no local addressing. The protocol is most naturally understood by describing the two-qubit system in the Bell basis:
\begin{eqnarray}
  \ket{\Psi^\pm} &=& \frac{1}{\sqrt{2}}(\ket{01} \pm \ket{10}) \\ 
  \ket{\Phi^\pm} &=& \frac{1}{\sqrt{2}}(\ket{00} \pm \ket{11}) 
\end{eqnarray}
We prepare the system in $\ket{00}$, and after a global $X(\pi/2)$ pulse, we prepare the state
\begin{eqnarray}
  \ket{\psi_1} &=& \frac{1}{2}(\ket{00} - i \ket{01} - i \ket{10} - \ket{11})
\end{eqnarray}
The controlled-phase gate creates the state
\begin{eqnarray}
  \ket{\psi_2} = \mathcal{CZ} \ket{\psi_1} &=& \frac{1}{2}(\ket{00} + i \ket{01} + i \ket{10} + \ket{11}) \\
			   &=& \frac{1}{\sqrt{2}} (\ket{\Phi^+} + i \ket{\Psi^+})
\end{eqnarray}
The states $\ket{\Phi^+}$ and $\ket{\Psi^+}$ are both within the triplet manifold of the two qubits and are coupled resonantly by a global drive field to form an effective two level system with twice the single-particle Rabi frequency. A $\pi/2$ pulse within this effective two-level system corresponds to a $\pi/4$ pulse at the single-particle Rabi frequency, and maps:
\begin{eqnarray}
  \ket{\psi_2} = \frac{1}{\sqrt{2}}(\ket{\Phi^+} + i \ket{\Psi^+}) \to \ket{\psi_3} = \ket{\Phi^+}
\end{eqnarray}

\begin{figure*}
  \includegraphics{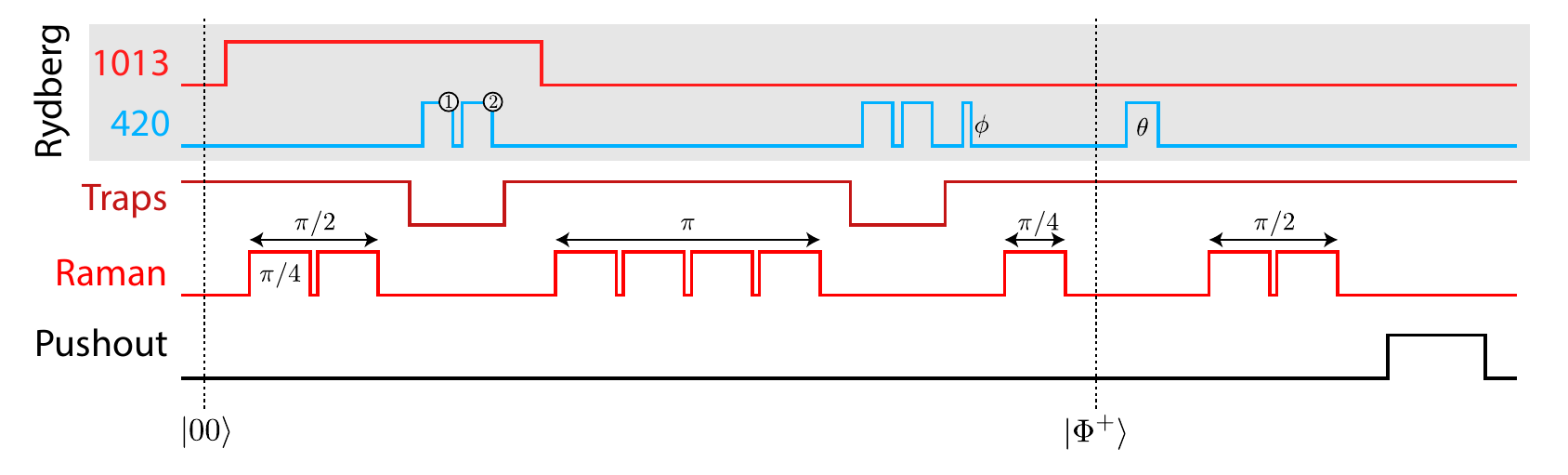}
  \caption{
	\textbf{Full pulse sequence for Bell state measurements.}
	The sequence begins with both atoms in $\ket{0}$ and a global $X(\pi/2)$ pulse (produced by two $\pi/4$ Raman pulses) to put both atoms in $\ket{-}_y$. Then, while the 1013 nm laser is on, the 420 nm laser is applied in two pulses (with a relative phase between the pulses) to enact the CZ gate, along with global phase shifts coming from the light shift of the 420 nm laser. A global $X(\pi)$ pulse flips the qubit states, at which point the same blue pulses are applied but now in the absence of 1013 nm light. This negates the effect of light shifts in the first portion of the CZ gate implementation. Then, to turn the CZ gate into the canonical $\mathcal{CZ}$ gate, we add a global phase by again pulsing the 420 nm laser and using the light shift to accumulate the appropriate phase correction. A subsequent global $X(\pi/4)$ pulse prepares the two atoms in the Bell state $\ket{\Phi^+}$. A final 420 nm laser pulse can be used to add dynamical phase to this Bell state, which can be detected by a subsequent global $X(\pi/2)$ for measuring parity oscillations. Finally, we push out atoms in $\ket{1}$ to detect populations.
  }
  \label{fig:sequence}
\end{figure*}

\section{Implementation of CCZ gate}
We implement the controlled-controlled-phase (CCZ) gate in the regime in which nearest neighbors are constrained by the Rydberg blockade, but next-nearest neighbors have only weak interactions. In light of this, the CCZ gate that we aim to implement is motivated by the fact that both edge atoms can simultaneously blockade the middle (target) atom. In particular, we consider the following scheme to implement CCZ that involves local excitation to Rydberg states:
\begin{enumerate}
  \item Apply a $\pi$ pulse on both edge atoms, transferring all of their population in $\ket{1}$ to $\ket{r}$.
  \item Apply a $2\pi$ pulse on the center atom, exciting from $\ket{1}$ to $\ket{r}$ and back to $\ket{1}$, accumulating a $\pi$ phase shift only if neither edge atom is blockading this central atom and the atom is in $\ket{1}$.
  \item Apply another $\pi$ pulse on the edge atoms to return any population from $\ket{r}$ to $\ket{1}$.
\end{enumerate}
  Such a protocol realizes the following unitary:
\begin{equation}
  \rm{CCZ} = 
  \begin{pmatrix}
	1 & 0 & 0 & 0 & 0 & 0 & 0 & 0 \\ 
	0 &-1 & 0 & 0 & 0 & 0 & 0 & 0 \\ 
	0 & 0 &-1 & 0 & 0 & 0 & 0 & 0 \\ 
	0 & 0 & 0 &-1 & 0 & 0 & 0 & 0 \\ 
	0 & 0 & 0 & 0 &-1 & 0 & 0 & 0 \\ 
	0 & 0 & 0 & 0 & 0 & 1 & 0 & 0 \\ 
	0 & 0 & 0 & 0 & 0 & 0 &-1 & 0 \\ 
	0 & 0 & 0 & 0 & 0 & 0 & 0 & 1 \\ 
  \end{pmatrix}
\end{equation}
This unitary is equivalent to the canonical controlled-controlled-phase gate, denoted $\mathcal{CCZ} = \mathds{1} - 2|111\rangle\langle 111|$ up to local rotations:
\begin{center}
\Qcircuit @C=0.8em @R=1.2em {
\\
& \multigate{2}{\mathcal{CCZ}}  & \qw &  &   & & \gate{X(\pi)}  & \multigate{2}{\rm{CCZ}}  & \gate{Z(\pi)} & \gate{X(\pi)} & \qw \\
& \ghost{\mathcal{CCZ}} 		& \qw &  & = & & \qw 		 	& \ghost{\rm{CCZ}}         & \qw		   & \qw 		   & \qw \\
& \ghost{\mathcal{CCZ}} 		& \qw &  &   & & \gate{X(\pi)}  & \ghost{\rm{CCZ}}         & \gate{Z(\pi)} & \gate{X(\pi)} & \qw \\
}
\end{center}

\begin{figure}
  \includegraphics{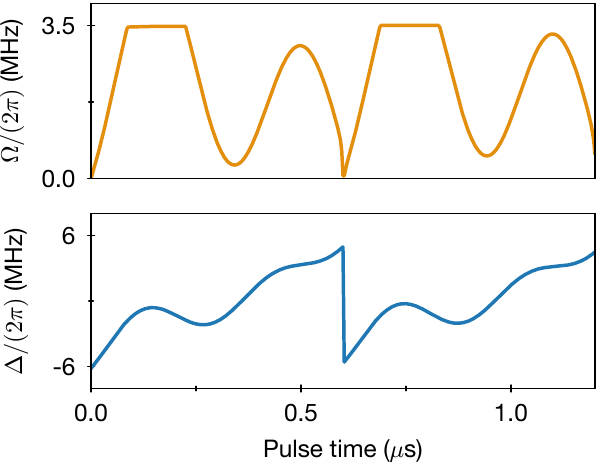}
  \caption{\textbf{Optimal control pulse for CCZ implementation.} Time variation of Rydberg Rabi frequency and detuning to approximately implement the CCZ gate with numerically simulated fidelity $97.6\%$.}
  \label{fig:opt_control}
\end{figure}

In the absence of local excitation to Rydberg states, we find that global Rydberg coupling can still approximately realize this unitary, although this implementation is limited by the finite next-nearest neighbor interaction. The global amplitude and frequency modulated pulse is found through the RedCRAB optimal control algorithm \cite{SIdCrab2015, SIREdCrab2018} and is shown in Fig.~\ref{fig:opt_control}. The optimized pulse has a duration of $1.2~\mu$s and achieves a numerically simulated gate fidelity of $97.6\%$.

\section{Echo procedure for CZ and CCZ}
The Rydberg pulse which implements the CZ or the CCZ gate includes both a 1013~nm laser field and a 420~nm laser field, the latter of which adds a differential light shift to the qubit levels of $\sim 2\pi \times 3~$MHz. To correct for the phase accumulated due to this light shift, after the CZ gate we apply a qubit $X(\pi)$ rotation on all atoms and then apply the same 420~nm pulse used for the CZ gate, but this time in the absence of 1013~nm light. 
The single particle phase $\phi$ (main text, eq.~(1)) inherent in the design of the CZ protocol is separately corrected by an additional short pulse of the 420~nm laser. The full detailed pulse sequence is shown in Fig.~\ref{fig:sequence}.

\section{State readout through atom loss}
Our primary technique for state readout is to apply a resonant laser pulse that heats atoms in $\ket{1}$ (in $F=2$, more generally) out of the tweezers, after which we take a fluorescence image of remaining atoms in $\ket{0}$. This method correctly identifies atoms in $\ket{0}$, but can mistake atoms that were lost through background loss processes or by residual Rydberg excitation for atoms in $\ket{1}$, leading to an overestimation of the population in $\ket{1}$. For any measurements involving Rydberg excitation, we therefore collect measurement statistics both with and without the pushout pulse, which provides an upper bound on how much leakage out of the qubit subspace occurred, and therefore also gives a lower bound on the true population in $\ket{1}$.

\begin{figure*}
  \includegraphics[width=0.7\textwidth]{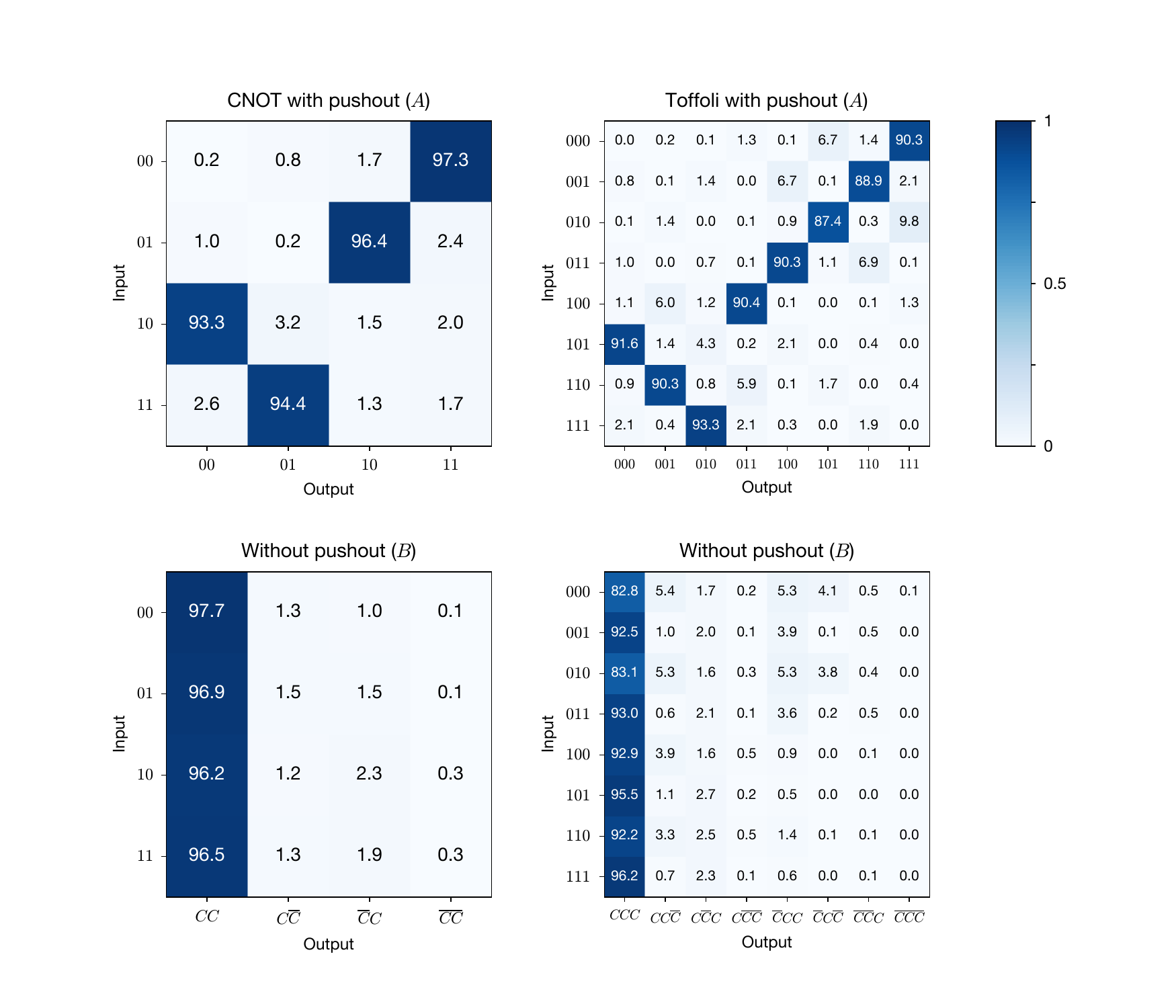}
  \caption{
	Here we show full measurement statistics for the CNOT and Toffoli truth tables. In both situations, for each input computational basis state, we measure the probability distribution of different output configurations both with and without the pushout pulse which removes $\ket{1}$ population, corresponding to the $A$ matrix and $B$ matrix, respectively. The output distribution of the $A$ matrix is mainly associated with qubit levels $\ket{0}$ and $\ket{1}$ according to whether the atom is present or absent. However, this approach overestimates population in $\ket{1}$ since leftover population in the Rydberg state and losses due to other processes lead to the same measurement outcome as $\ket{1}$. To distinguish this effect, we measure without the pushout pulse (bottom row) to assess how much population is left in the computational subspace ($C$), rather than lost into the Rydberg state and therefore out of the computational subspace ($\overline{C}$). Comparing these two measurements provides a lower bound on the true atomic populations in the $\ket{0}$ and $\ket{1}$ qubit states.
  }
  \label{figure:FullMeasurement}
\end{figure*}

We illustrate this procedure in the context of two-qubit experiments. Let us denote the two types of measurements as $A$ (in which we apply the pushout of $\ket{1}$ atoms) and $B$ (in which we disable the pushout). For each measurement procedure, we obtain statistics of observing the four two-qubit states, consisting of `lost' or `present' for each qubit. The $A$ vector associates these as $\ket{0}$ and $\ket{1}$, so $A_{ij}$ (for $i,j \in \{0,1\}$) denotes the probability of identifying the left and right atom in $0,1$ through the simple loss/presence analysis. However, the atoms can be not only in the qubit states $0,1$ but they can also be lost from the trap or in the Rydberg state, which in both cases will be detected as `lost'. Let us denote $C$ as the computational subspace containing $\ket{0}$ and $\ket{1}$, and denote $\overline{C}$ as anything outside this subspace (including Rydberg population or loss). The $B$ vector measures whether the atoms are in $C$ (either $\ket{0}$ or $\ket{1}$), or not ($\overline{C}$), so is denoted $B_{ij}$ where $i, j \in \{C, \overline{C}\}$.

Both $A_{ij}$ and $B_{ij}$ can be explicitly expressed in terms of the underlying atomic populations $p_{\alpha \beta}$, where $\alpha, \beta \in \{0, 1, \overline{C}\}$, as follows;

\begin{eqnarray}
  A_{00} &=& p_{00} \\
  A_{01} &=& p_{01} + p_{0\overline{C}}\\
  A_{10} &=& p_{10} + p_{\overline{C}0} \\
  A_{11} &=& p_{11} + p_{1\overline{C}} + p_{\overline{C}1} + p_{\overline{C}\overline{C}}
\end{eqnarray}

\begin{eqnarray}
  B_{CC} &=& p_{00} + p_{01} + p_{10} + p_{11} \\
  B_{C\overline{C}} &=& p_{0\overline{C}} + p_{1\overline{C}} \\
  B_{\overline{C}C} &=& p_{\overline{C}0} + p_{\overline{C}1} \\
  B_{\overline{C}\overline{C}} &=& p_{\overline{C}\overline{C}}
\end{eqnarray}

Measuring $A_{ij}$ and $B_{ij}$, we can now solve for the atomic populations of interest: $p_{00}, p_{01}, p_{10}, p_{11}$.

\begin{eqnarray}
  p_{00} &=& A_{00} \\
  p_{01} &=& A_{01} - B_{C\overline{C}} + p_{1\overline{C}} \\
  p_{10} &=& A_{10} - B_{\overline{C}C} + p_{\overline{C}1} \\
  p_{11} &=& A_{11} - B_{C\overline{C}} - B_{\overline{C}C} - B_{\overline{C}\overline{C}}  + (p_{0\overline{C}} + p_{\overline{C}0})
\end{eqnarray}

Since all probabilities are non-negative and $B_{C\overline{C}} + B_{\overline{C}C} + B_{\overline{C}\overline{C}} = 1-B_{CC}$, we have our lower bounds for the true populations:
\begin{eqnarray}
  p_{00} &=& A_{00} \\
  p_{01} &\ge& A_{01} - B_{C \overline{C}} \\
  p_{10} &\ge& A_{10} - B_{\overline{C}C} \\
  p_{11} &\ge& A_{11} - (1 - B_{CC})
\end{eqnarray}

This is the analysis carried out for the Bell state populations, the CNOT truth table, and the Toffoli truth table (extended to three qubits). For the truth tables, the analysis is carried out for each measurement configuration (corresponding to a different input computational basis state) separately, shown as the rows in the matrices of Fig.~\ref{figure:FullMeasurement}.

\section{Correcting for state preparation and measurement errors}
We consider the problem of correcting a measured fidelity for state preparation and measurement (SPAM) errors. We denote $P$ as the probability to correctly initialize and measure all qubits; generally, $P = (1 - \epsilon)^N$ for single-particle SPAM error rate of $\epsilon$. The measured fidelity is related to the `corrected fidelity' according to:
\begin{equation}
  \mathcal{F} = P \times \mathcal{F}^{\rm{c}} + (1-P) \times \mathcal{F}^{\rm{false}}
  \label{eq:SPAM}
\end{equation}
Here $\mathcal{F}^{\rm{false}}$ denotes the false contribution to the measured fidelity signal in cases in which SPAM errors occur. The main subtlety in performing this correction is properly evaluating the potential false contribution $\mathcal{F}^{\rm{false}}$.

Experimentally, the SPAM error is $\epsilon = 1.2(1)\%$ per qubit, consisting of two effects: first, the optical pumping into $\ket{0}$ has an error probability of $0.7(1)\%$, constituting a state preparation error. Second, there is a small chance that an atom can be lost due to a background collision either before or after the Bell state circuit is performed. Loss before the circuit contributes as a state preparation error; loss after the circuit but prior to the readout fluorescence image contributes as a measurement error. The total background loss contribution is $0.5(1)\%$ error per atom.

\subsection{Bell state fidelity}
The total probability that no errors occur on either of two qubits is $P = 97.6(2)\%$. Equation \eqref{eq:SPAM} holds for both the population measurement and the parity oscillation measurement separately. The population measurement explicitly only counts lower bounds on the population of atoms within the qubit subspace (see section: ``State readout through atom loss''). Therefore, in cases where an atom is lost there is no false contribution to the measured fidelity. However, our measured fidelity does not distinguish between atoms pumped into magnetic sublevels outside of the qubit subspace. We estimate that in cases when one of the two atoms are prepared in an incorrect magnetic sublevel ($1.4(2)\%$ probability), there can be a false contribution of $\mathcal{F}^{\rm{false}} = 1 - \cos^2(7\pi/8) \approx 15\%$ (calculated by evaluating the quantum circuit in the main text Fig.~3a with one atom not participating). The lower bound on the measured probablilities $p_{00} + p_{11} \ge 95.8(3)\%$ therefore set a lower bound on the corrected populations: $p_{00}^{\rm{c}} + p_{11}^{\rm{c}} \ge 97.9(4)\%$.

On the other hand, the parity oscillation amplitude receives no false contribution from cases when an atom is prepared in the wrong sublevel or is lost, because this error is independent of the accumulated phase and therefore does not oscillate as a function of the phase accumulation time. The false contribution is therefore $\mathcal{F}^{\rm{false}} = 0$. In this case, the coherence $C$ (given by the amplitude of the parity oscillation) is related to the corrected coherence by $C = P \times C^{\rm{c}}$. Since $C = 94.2(4)\%$, we obtain a corrected coherence of $C^{\rm{c}} = 96.5(4)\%$. The total SPAM-corrected Bell state fidelity, then, is $\mathcal{F}^{\rm{c}} = \frac{1}{2}(p_{00}^{\rm{c}} + p_{11}^{\rm{c}} + C^{\rm{c}}) \ge 97.2(3)\%$.

\subsection{CNOT Truth Table}
We measure the truth table by performing the CNOT gate on each computational basis state. The basis states are prepared with finite fidelity, as measured and shown in the main text Fig. 3e. For each basis state, we wish to assess how the finite output fidelity in the target state compares to the finite initialization fidelity to determine how well the gate performs on this input state. We establish a probability $P_{ij}$ of no SPAM error occurring for each measurement setting (where $ij$ denotes the setting in which we initialize the computational basis state $\ket{ij}$). Additionally, we measure a lower bound on the output probability in the target state, $\mathcal{F}_{ij}$. 

We now consider false contributions to the measured fidelity. When an error involving atom loss occurs, there is no false contribution to fidelity since fidelity only measures atom population within the qubit subspace. Alternatively, in cases when the wrong computational basis state is prepared, then $\mathcal{F}^{\rm{false}}$ is bounded above by the largest unwanted element of the truth table, or $<4\%$. The total false contribution therefore is $(1-P) \times \mathcal{F}^{\rm{false}} <(3\%) \times (4\%) \lesssim 0.1\%$. This contribution is below our measurement resolution and we do not account for it. The corrected fidelity is therefore just given by $\mathcal{F}^{\rm{c}}_{ij} = \mathcal{F}_{ij}^{\rm{meas}} / P_{ij}$.
The average corrected truth table fidelity, given by the average of $\mathcal{F}^{\rm{c}}_{ij}$,  is therefore $\mathcal{F}^{\rm{c}}_{\rm{CNOT}} \ge 96.5\%$ (see Table \ref{table:measurements}).

\subsection{Toffoli Truth Table}
We perform the same analysis to evaluate the corrected Toffoli truth table fidelity as for the CNOT truth table.
The average corrected truth table fidelity is $\mathcal{F}_{\rm{Toff}}^{\rm{c}} \ge 87.0\%$ (see Table \ref{table:measurements}).

\begin{table}
  \begin{center}
	\begin{tabular}{r | c | c | c }
	  & Raw outcomes & Lower bound & Corrected \\
	  \hhline{|=|=|=|=|}
	  Bell state populations & $97.6\%$ & $95.8\%$ & $97.9\%$  \\
	  Bell state coherences & $94.2\%$ & $94.2\%$ & $96.5\%$  \\
	  \hline
	  \textbf{Bell state fidelity} & $95.9\%$ & $95.0\%$ & $97.2\%$  \\
	  \hhline{|=|=|=|=|}
	  CNOT: Input 00 & $97.3\%$ & $95.0\%$ & $96.5\%$ \\
	  01 & $96.4\%$ & $94.9\%$ & $97.9\%$ \\
	  10 & $93.3\%$ & $93.3\%$ & $96.3\%$ \\
	  11 & $94.4\%$ & $93.1\%$ & $95.4\%$ \\
	  \hline
	  \textbf{CNOT Truth table} & $95.4\%$ & $94.1\%$ & $96.5\%$ \\
	  \hhline{|=|=|=|=|}
	  Toffoli: Input 000 & $90.3\%$ & $73.1\%$ & $75.1\%$ \\
	  				 001 & $88.9\%$ & $82.6\%$ & $86.2\%$ \\
	  				 010 & $87.4\%$ & $73.0\%$ & $76.0\%$ \\
	  				 011 & $90.3\%$ & $86.7\%$ & $90.0\%$ \\
	  				 100 & $90.4\%$ & $84.3\%$ & $87.4\%$ \\
	  				 101 & $91.6\%$ & $91.6\%$ & $95.7\%$ \\
	  				 110 & $90.3\%$ & $87.0\%$ & $90.5\%$ \\
	  				 111 & $93.3\%$ & $91.0\%$ & $95.0\%$ \\
	  \hline
	  \textbf{Toffoli Truth table} & $90.3\%$ & $83.7\%$ & $87.0\%$ \\
	\end{tabular}
  \end{center}
  \caption{Summary of measurement results. Raw outcomes correspond to simple assignment of atom presence to qubit state 0 or 1. The lower bound comes from subtracting a conservative upper bound estimate on how much leakage out of the qubit subspace there may be, as determined by a separate measurement in which we do not push out $\ket{1}$ atoms. The corrected column shows the fidelities corrected for SPAM errors.}
  \label{table:measurements}
\end{table}

\section{Limited tomography of Toffoli gate}
\begin{figure}
  \includegraphics{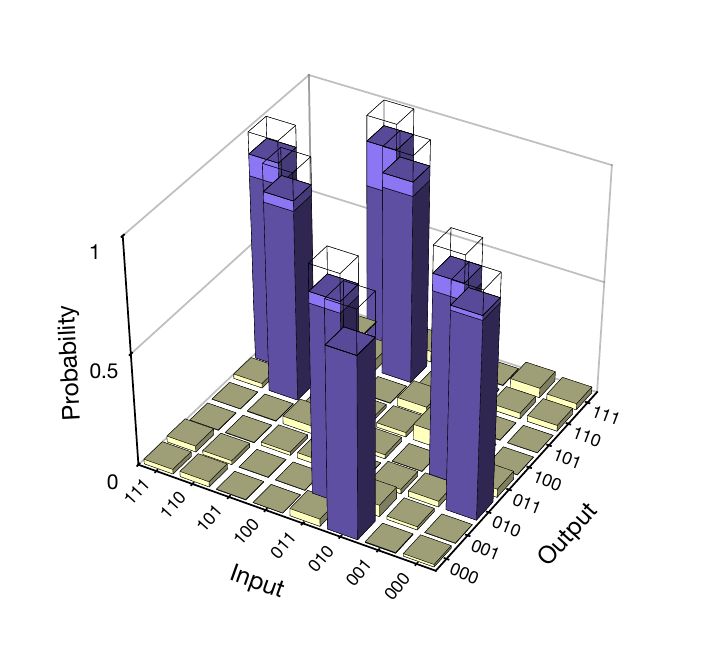}
  \caption{\textbf{Limited tomography of Toffoli gate.} The raw target probabilities average to  $88.0(3)\%$. Since four of the measurement configurations are precisely global $X(\pi)$ gates applied to the other four input states, we can compare these output distributions to properly account for leftover Rydberg population, similar to the procedure discussed in \emph{State readout through atom loss}. We establish the limited tomography fidelity is therefore $\mathcal{F} \ge 81.5(5)\%$. Corrected for SPAM errors, the fidelity is $\ge 86.2(6)\%$.
  }
  \label{fig:LimitedTomography}
\end{figure}
The truth table of the Toffoli gate provides a representation of the magnitude of the matrix elements of the gate expressed in the logical basis. However, the measured populations carry no information about the relative phases between the different entries. Performing a similar procedure as the truth table but rotating the Toffoli gate to act on the $X$-basis instead of the $Z$-basis, makes it possible to recover some information about these phases. A restricted version of such a procedure has been used before as a way to characterize the fidelity of the Toffoli gate \cite{SIMonroeGrover2017}, and has been dubbed ``Limited Tomography''. The procedure consists of initializing all the computational basis states in the Z-basis, and then applying an $X(\pm \pi/2)$ rotation to all qubits before and after a Toffoli gate. The sign is chosen to be $X(+\pi/2)$ when the target qubit is initialized in $\ket{0}$ and $X(-\pi/2)$ when the target qubit is initialized in $\ket{1}$.
\centerline{\Qcircuit @C=0.8em @R=1.2em {
\\
\lstick{\ket{q_{0}}} & \gate{X(\pm \pi/2)} & \ctrl{1} & \gate{X(\pm \pi/2)} & \meter \\
\lstick{\ket{q_{1}}} & \gate{X(\pm \pi/2)} & \targ & \gate{X(\pm \pi/2)} & \meter \\
\lstick{\ket{q_{2}}} & \gate{X(\pm \pi/2)} & \ctrl{-1} & \gate{X(\pm \pi/2)} & \meter \\
\\
}}
Conditioning the sign of the rotation on the state of the target qubit enforces that the target qubit is always in the same state $\ket{+}_y$ prior to the action of the Toffoli gate itself. 

The Toffoli gate implemented in our system, which includes an echo pulse that acts as a global $X(\pi)$ gate (see main text, Fig. 4), is described ideally by the unitary matrix:

\begin{equation}
 \label{eq:ToffoliIdeal}
  T_{\text{Ideal}}=
  \begin{pmatrix}
    0 & 0 & 0 & 0 & 0 & i & 0 & 0 \\
    0 & 0 & 0 & 0 & 0 & 0 & 1 & 0 \\
    0 & 0 & 0 & 0 & 0 & 0 & 0 & -i \\
    0 & 0 & 0 & 0 & 1 & 0 & 0 & 0 \\
    0 & 0 & 0 & 1 & 0 & 0 & 0 & 0 \\
    0 & 0 & -1 & 0 & 0 & 0 & 0 & 0 \\
    0 & 1 & 0 & 0 & 0 & 0 & 0 & 0 \\
    -1 & 0 & 0 & 0 & 0 & 0 & 0 & 0 \\
  \end{pmatrix},
\end{equation}

Performing the limited tomography procedure on this unitary should result in the following output truth table:
\begin{equation}
 \label{eq:LTIdeal}
 \rm{Lim}[T_{\text{Ideal}}]=
  \begin{bmatrix}
    0 & 0 & 1 & 0 & 0 & 0 & 0 & 0 \\
    0 & 0 & 0 & 1 & 0 & 0 & 0 & 0 \\
    1 & 0 & 0 & 0 & 0 & 0 & 0 & 0 \\
    0 & 1 & 0 & 0 & 0 & 0 & 0 & 0 \\
    0 & 0 & 0 & 0 & 0 & 0 & 1 & 0 \\
    0 & 0 & 0 & 0 & 0 & 0 & 0 & 1 \\
    0 & 0 & 0 & 0 & 1 & 0 & 0 & 0 \\
    0 & 0 & 0 & 0 & 0 & 1 & 0 & 0 \\
  \end{bmatrix},
\end{equation}
where each row shows the target output probabilities for a given input state. However, if the Toffoli gate is allowed to deviate from the ideal unitary by arbitrary phases $\phi_j$ according to
\begin{equation}
 \label{eq:ToffoliPhases}
  T_{\phi}=
  \begin{pmatrix}
	0 & 0 & 0 & 0 & 0 & ie^{i\phi_1} & 0 & 0 \\
	0 & 0 & 0 & 0 & 0 & 0 & e^{i\phi_2} & 0 \\
	0 & 0 & 0 & 0 & 0 & 0 & 0 & -ie^{i\phi_3} \\
	0 & 0 & 0 & 0 & e^{i\phi_4} & 0 & 0 & 0 \\
	0 & 0 & 0 & e^{i\phi_5} & 0 & 0 & 0 & 0 \\
	0 & 0 & -e^{i\phi_6} & 0 & 0 & 0 & 0 & 0 \\
	0 & e^{i\phi_7} & 0 & 0 & 0 & 0 & 0 & 0 \\
	-e^{i\phi_8} & 0 & 0 & 0 & 0 & 0 & 0 & 0 \\
  \end{pmatrix},
\end{equation}
then the limited tomography truth table reflects this phase deviation. In particular, each truth table matrix element in which the limited tomography should produce unity will instead result in a peak probability of $|\frac{1}{8} \sum_{j} e^{i\phi_j}|^2$. The average fidelity of the limited tomography truth table therefore reflects how close the phases on the Toffoli unitary are to their ideal values, and can only reach unity if each phase is correct. Our measured limited tomograpy truth table is shown in Fig.~\ref{fig:LimitedTomography}.

It is worth noting that the limited tomography protocol only makes use of four of the 8 $X$-basis input states, as seen by the fact that the target qubit is always initialized in $\ket{+}$. This makes four out of the eight measurements equivalent to the other four up to a global $X(\pi)$ rotation at the end. Comparing these two sets of measurements gives a constraint on the probability of leakage out of the qubit subspace, similarly to the approach described in the section ``State readout through atom loss.''

\begin{figure}
  \includegraphics{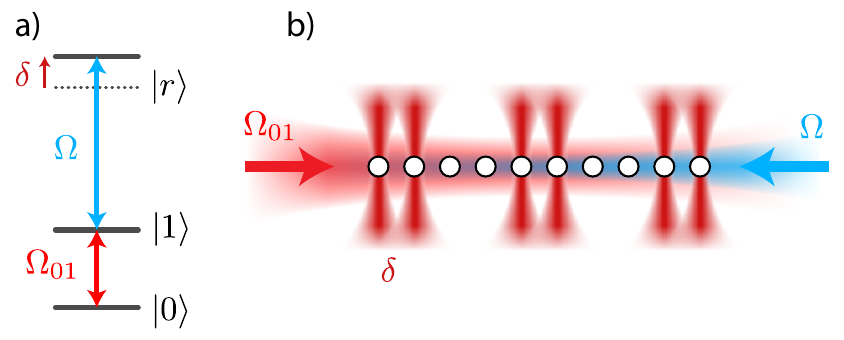}

  \caption{Parallel gate implementation in a contiguous chain of atoms.
  (a) Local addressing lasers can shift the frequency of the Rydberg transition from $\ket{1}$ to $\ket{r}$ by $\delta$ without changing the $\ket{0} \leftrightarrow \ket{1}$ frequency.
  (b) The local addressing lasers are focused onto a subset of qubits on which we aim to perform parallel multi-qubit gates. The global Rydberg coupling laser is tuned to the light-shifted resonance, so that only the locally addressed atoms are coupled to the Rydberg state for gate implementation.}
  \label{fig:Contiguous}
\end{figure}

\section{Parallel gate implementation in a contiguous array}
The experiments performed here involve parallel multi-qubit gate implementation on separated pairs of atoms, where the inter-pair interaction is negligible. However, one can extend this protocol to parallel gate implementation in a contiguous chain of atoms, as illustrated in Fig.~\ref{fig:Contiguous}. We consider an additional local addressing laser system which can address an arbitrary subset of atoms, using for example an acousto-optic deflector. Specifically, one can select a wavelength for this laser such that the imparted light shift affects the $\ket{0}$ and $\ket{1}$ states equally, but differently from the Rydberg state $\ket{r}$. In such a case, the light shift from this new local addressing laser does not apply any qubit manipulations, but instead simply shifts the effective Rydberg resonance. Near-infrared wavelengths tuned far from any ground state optical transition ($\lambda \gtrsim  820~$nm) are suitable for Rubidium.

With such a system, we could illuminate all pairs of adjacent atoms on which we intend to perform two-qubit gates, and then by tuning the Rydberg laser to the light-shifted resonance we would perform the multi-qubit gate on all pairs in parallel. The only constraint is that there must be sufficient space between addressed pairs such that the interaction (cross-talk) between them is negligible  in a particular layer of gate implementation.


\end{document}